\documentclass[a4paper,11pt]{article}
\usepackage{pos}

\usepackage{tikz}
\usetikzlibrary{shapes,arrows.meta,positioning, decorations.markings}

\usepackage{enumitem}
\usepackage{xspace}

\newcommand{\trace}[1]{\left\langle #1 \right\rangle}
\newcommand{\minibar}{\texttt{MINIBAR}\xspace}

\title{The anomalous Lagrangian in ChPT at NNLO}

\author[a]{Johan Bijnens}
\author[a,b]{Nils Hermansson-Truedsson}
\author*[a]{Joan Ruiz-Vidal}

\affiliation[a]{Division of Particle and Nuclear Physics, Department of Physics, Lund University,\\
Box 118, SE 221 00 Lund, Sweden}

\affiliation[b]{Higgs Centre for Theoretical Physics, School of Physics and Astronomy, The University of Edinburgh,\\ James Clerk Maxwell Building,
Peter Guthrie Tait Road,
Edinburgh,
EH9 3FD}

\emailAdd{johan.bijnens@fysik.lu.se}
\emailAdd{nils.hermansson-truedsson@ed.ac.uk}
\emailAdd{joan.ruiz-vidal@fysik.lu.se}

\abstract{The anomalous Lagrangian in mesonic Chiral Perturbation Theory, of odd intrinsic parity, is determined to next-to-next-to-leading order thereby completing the order $p^8$ Lagrangian. A schematic view of its construction with the \minibar package for Mathematica is presented and the final operator count is discussed for a general number of light quark flavours as well as for the physical cases $N_f=2,3$. The number of operators in our explicit construction of the Lagrangian basis is consistent with the number derived using the Hilbert series in the literature.
}

\FullConference{The European Physical Society Conference on High Energy Physics (EPS-HEP2023)\\
 21-25 August 2023\\
Hamburg, Germany\\}

\begin{document}
\maketitle

\section{Introduction}
\vspace{-0.4cm}
An anomaly occurs when a symmetry is respected in the bare Lagrangian, at the classical level, but the corresponding current is not conserved due to quantum effects.
In the SM, anomalies first arise in triangle diagrams with fermions running in the loops, although the summation over three colours and fermion and antifermion cancels them for the gauge currents. In the low-energy effective theory of quantum chromodynamics (QCD), Chiral Perturbation Theory (ChPT)~\cite{Weinberg:1978kz,Gasser:1983yg,Gasser:1984gg}\footnote{Find introductions to ChPT and all the standard definitions in \textit{e.g.} Refs.~\cite{Pich:2018ltt, Scherer:2012xha}. For a short summary, see Section 2.1 of our main article~\cite{Bijnens:2023hyv}.}, the underlying anomalies are reproduced already at tree level, as there are Lagrangian terms that do not respect chiral symmetry, with direct phenomenological implications for example in the (classically forbidden) decay $\pi^0 \to \gamma \gamma$. In the perturbative expansion of the ChPT Lagrangian, in powers of the momentum transfer of the process $p^{2n}$, the anomaly itself is contained in the Wess-Zumino-Witten (WZW) term, at $\mathcal{O}(p^4)$~\cite{WESS197195,Witten:1983tw}. In four dimensions this is in fact an infinite series of interaction terms that nevertheless depend only on one parameter, which is associated to the number of colours $N_C$ of QCD. This implies that there are no unknown low-energy constants (LECs) associated to anomalous processes to first order. 
Another characteristic of the WZW term is that it is of odd intrinsic parity. Intrinsic parity transformations act on the fields like normal parity but without reversing the sign of the space coordinates. Pseudoscalar meson fields transform as $\phi(\vec{x},t) \to -\phi(\vec{x},t)$ and therefore these interactions only contribute to processes with an odd number of mesons (and axial vectors). As already pointed out, all anomalous interactions that do not respect chiral symmetry are contained in the WZW term of $\mathcal{O}(p^4)$. However, at higher orders one can build effective operators that are also of odd intrinsic parity, albeit perfectly chirally symmetric. This broader class of Lagrangian operators, characterized by the odd intrinsic parity, is collectively referred to as the anomalous sector. Beyond $p^4$, this sector does contain a number of LECs that have to be determined from experiments or lattice calculations. 
The objective of this work is to determine explicitly the next-to-next-to-leading (NNLO) anomalous Lagrangian of ChPT. Together with the lower orders, $p^4$~\cite{WESS197195,Witten:1983tw} and $p^6$~\cite{Bijnens:2001bb,Ebertshauser:2001nj}, and the non-anomalous (even intrinsic parity) sector of order $p^2$--$p^8$~\cite{Gasser:1983yg,Gasser:1984gg,PhysRevD.53.315,Bijnens:1999sh,Bijnens:2018lez}, our result~\cite{Bijnens:2023hyv} completes the ChPT Lagrangian through $\mathcal{O}(p^8)$. For future phenomenological applications, the renormalization at $\mathcal{O}(p^8)$ for both the anomalous and non-anomalous sectors also has to be studied~\cite{Hermansson-Truedsson:2020rtj}.

\vspace{-0.3cm}
\section{What to do}
\vspace{-0.4cm}
\newcommand{\cP}{$\mathcal{P}$}
\newcommand{\cC}{$\mathcal{C}$}
\newcommand{\cH}{$\mathcal{H}$}
\newcommand{\ind}{{\mu\nu\rho\sigma}}
Let us see what is the shape of the Lagrangian operators that we must construct. As the effective theory of QCD at low energies, the ChPT Lagrangian must respect its symmetries: hermitian conjugation ($\mathcal{H}$), charge conjugation ($\mathcal{C}$), and parity ($\mathcal{P}$). Under $\mathcal{P}$, the components of any tensor $X^{\mu\nu}$ transform as $\delta(\mu)\delta(\nu) X_{\mathcal{P}}^{\mu\nu}$, with $\delta(0) = - \delta(\mu=1,2,3) = 1$ due to the space inversion in four dimensions. In Lorentz invariant structures with these tensors, the factors disappear since $\delta(\mu)^2 = 1$. However, when four indices are contracted with a Levi-Civita symbol, we have that 
\begin{equation}
    X^\ind \varepsilon^{\ind} \stackrel{\mathcal{P}}{\longrightarrow} X_{\mathcal{P}}^{\ind}\delta(\mu)\delta(\nu)\delta(\rho)\delta(\sigma)\varepsilon^{\mu \nu \rho \sigma} = - X_{\mathcal{P}}^{\ind} \varepsilon^{\mu \nu \rho \sigma} \, 
\end{equation}
where we used that the four indices must be different in $\varepsilon^{\ind}$. This \textit{external} sign must be compensated by the rest of the operator, which must be of odd intrinsic parity $X_{\mathcal{P}}^{\ind} = - X^{\ind}$. If all the fields composing the operator, \textit{e.g.} $\Psi^{\mu \nu}$, transform into themselves as $\Psi_{\mathcal{P}}^{\mu\nu} = \pm\Psi^{\mu\nu}$, it is enough to select the operators with an odd number of \cP-odd fields. This is one of the advantages of the field basis that we will use, the $u_\mu$ basis, whose transformations under discrete symmetries are defined in Table 2 of the main article~\cite{Bijnens:2023hyv}. 
Under the chiral group $SU(N_f)_L \times SU(N_f)_R$, where $N_f$ is the number of light quark flavours, these $N_f \times N_f$ matrices transform linearly with the elements of the unbroken subgroup $h \in SU(N_f)_V$ as $X \to h X h^\dagger$. This is an additional advantage of the $u_\mu$ field parametrization, as all the flavour matrices transform in the same way and their product is automatically chirally invariant, following from $h^\dagger h = 1$
\footnote{We will also use a different parametrization of the fields, the left-right (LR) basis. This is done to obtain explicitly the contact operators with only external fields, i.e.~fields that are not the lightest pseudoscalar mesons acting as pseudo-Goldstone bosons of the chiral symmetry breaking, required for renormalization of the theory.}.

In practical terms, our objective is to find the most general set of linearly-independent operators $\{\mathcal{O}_{i}^{\mu\nu\rho\sigma}\}$, that have chiral order $p^8$, and are contracted to a Levi-Civita symbol, forming the anomalous Lagrangian
\begin{align}\label{eq:lagdef2n}
 \mathcal{L}_{8}^{\textrm{odd}} = 
 \sum _{i=1}^{N_{8}} a_{i} \, \mathcal{O}_{i}^{\mu\nu\rho\sigma} ~\varepsilon_{\mu\nu\rho\sigma} \, .
\end{align}
Here the coefficients $a_i$ are the (unknown) LECs. To achieve this, we must build all possible structures that respect the symmetries of our theory using combinations of the flavour matrices $\left\lbrace u_\mu,~f_{\pm~\mu\nu},~\chi_{\pm}\right\rbrace$ as well as their covariant derivatives $\nabla_\mu X$, and remove the redundant operators.

\vspace{-0.3cm}
\section{How to do it}
\vspace{-0.4cm}
Given the large number of possible combinations at order $p^8$ ($\sim244$k), this analytical calculation cannot be done by hand and every step needs to be fully programmed. For this, we developed the \textsc{Mathematica} package \minibar~\cite{MINIBAR}, that offers a set of configurable functions to build and reduce the operator set to a minimal basis in a time-efficient way. It also provides a simple interface to some of the methods of \textsc{SuiteSparse}~\cite{SPQR}, based on C++, to decompose large sparse matrices and thereby determine a minimal set of linearly-independent operators. An independent calculation was done in parallel using the compilable software for analytic calculations \textsc{Form}~\cite{Vermaseren:2000nd,Ruijl:2017dtg}.

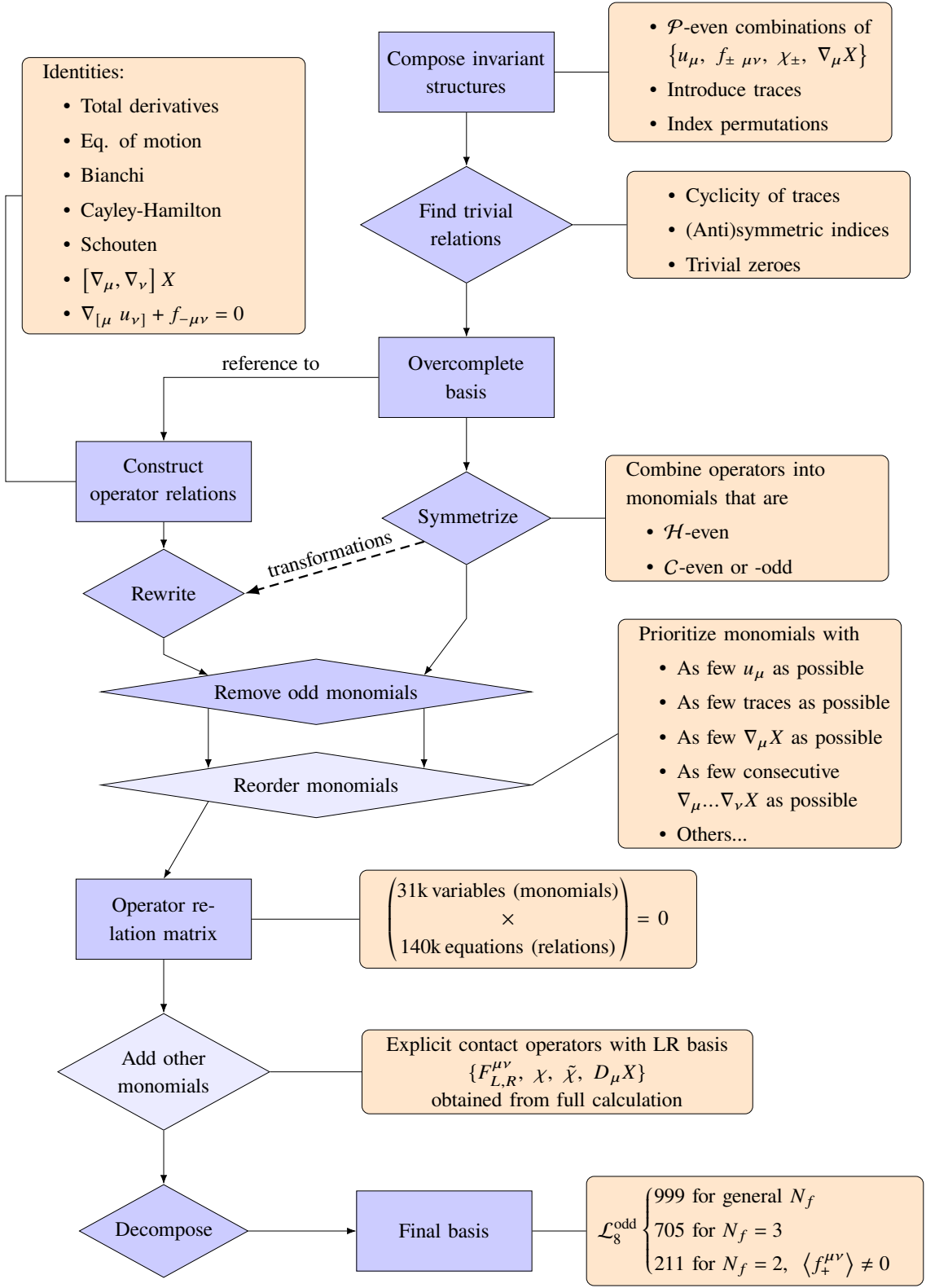
\begin{figure}
	\centering
	\resizebox{!}{0.93\textheight}{
		\begin{tikzpicture}
			\tikzstyle{stepbox} = [draw, fill=blue!20, text width=3cm, text centered, minimum height=1.5cm, rectangle]
			\tikzstyle{configbox} = [draw, fill=orange!20, text width=5cm, text centered, minimum height=1.5cm, rounded corners]
			\tikzstyle{actionbox} = [diamond, draw, fill=blue!20, text width=1.8cm, text centered, minimum height=1.5cm, shape aspect=1.8]
			\tikzstyle{optional} = [diamond, draw, fill=blue!8, text width=1.8cm, text centered, minimum height=1.5cm, shape aspect=1.8]
			\tikzstyle{arrow} = [-{Latex},,>=stealth]
            \tikzset{
  discontinuous arrow/.style={
    dash pattern=on 5pt off 3pt,
    -{Latex}, 
    line width=1pt
  }
}
			
			\node (compose) [stepbox] at (0,0) {Compose invariant structures};
			\node (trivial) [actionbox, below=1cm of compose] {Find trivial relations};
            \node (overcomplete) [stepbox, below=1cm of trivial] {Overcomplete basis};
			\node (symmetrize) [actionbox, below=1cm of overcomplete] {Symmetrize};

			\node (removeodd) [actionbox, text width=4cm, minimum width=8cm, minimum height=0.5cm, shape aspect=6, below left=2.5cm and 0.0 cm of symmetrize] {Remove odd monomials};
			\node (preferred) [optional, text width=4cm, minimum width=8cm, minimum height=0.5cm, shape aspect=6, below=0.5cm of removeodd] {Reorder monomials};
			
			\node (oprel) [stepbox, below left=3.5cm and 3cm of trivial] {Construct operator relations};
			\node (rewrite) [actionbox, below=0.5cm of oprel] {Rewrite};
			\node (relmat) [stepbox, below=4.5cm of rewrite] {Operator relation matrix};

			\node (extrarel) [optional, below=1cm of relmat] {Add other monomials};
			\node (decompose) [actionbox, below=1cm of extrarel] {Decompose};
			\node (independent) [stepbox, right=2cm of decompose] {Final basis};

			\node (system) [configbox, text width=6cm, right=2cm of relmat] {$\begin{pmatrix}
					31{\rm k} \, \text{variables (monomials)} \\
					\times \\
					140{\rm k} \, \text{equations (relations)}
				\end{pmatrix} = 0$};
			\node (contact) [configbox, text width=7cm, right=1.7cm of extrarel] {Explicit contact operators with LR basis\\ $\{ F_{L,R}^{\mu\nu} , ~\chi,~ \tilde\chi, ~D_\mu X\}$\\ obtained from full calculation};
			\node (result) [configbox, text width=6cm,  right=1cm of independent] {$\mathcal{L}_8^\text{odd} 
				\begin{cases}
					\begin{aligned}
						&\text{999 for general } N_f \\
						&\text{705 for } N_f = 3 \\
						&\text{211 for } N_f = 2,~\trace{f_{+}^{\mu \nu}}\neq 0
					\end{aligned}
				\end{cases}$};

			\node (composeConfig) [configbox, right=1cm of compose] {
				~\vspace*{-.4cm}\begin{itemize}[itemsep=-1pt]
					\item \cP-even combinations of $\left\lbrace u_\mu,~f_{\pm~\mu\nu},~\chi_{\pm},~\nabla_\mu X\right\rbrace$
					\item Introduce traces
					\item Index permutations
				\end{itemize}
			};
			\node (trivialConfig) [configbox, right=1cm of trivial] {~\vspace*{-.4cm}\begin{itemize}[itemsep=-1pt]
					\item Cyclicity of traces
					\item (Anti)symmetric indices
					\item Trivial zeroes
			\end{itemize}};
			\node (symmetrizeConfig) [configbox, align=left, right=1cm of symmetrize] {~~Combine operators into\\ ~~monomials that are \vspace*{-.2cm} \begin{itemize}[itemsep=-1pt]
					\item $\mathcal{H}$-even 
					\item $\mathcal{C}$-even or -odd
			\end{itemize}};
			\node (preferredConfig) [configbox, align=left, above right=-1.5cm and 3.6cm of preferred]    {~\hspace*{0.0cm} Prioritize monomials with \vspace*{-.2cm}    \begin{itemize}[itemsep=-1pt]
					\item As few $u_\mu$ as possible
					\item As few traces as possible
					\item As few $\nabla_\mu X$ as possible
					\item As few consecutive $\nabla_\mu ... \nabla_\nu X$ as possible
					\item Others...
			\end{itemize}};

			\node (oprelConfig) [configbox, align=left, above=2cm of oprel] {~~Identities: \vspace*{-0,2cm} \begin{itemize}[itemsep=-1pt]
					\item Total derivatives
					\item Eq. of motion
					\item Bianchi
					\item Cayley-Hamilton
					\item Schouten
					\item $\left[ \nabla_\mu, \nabla_\nu \right]X$
					\item $\nabla_{[\mu}~ u_{\nu]}+f_{-\mu \nu} = 0$
			\end{itemize}};

			\draw (compose.east) -- (composeConfig.west);
			\draw (trivial.east) -- (trivialConfig.west);
			\draw (symmetrize.east) -- (symmetrizeConfig.west);
			\draw (preferred.east) -- (preferredConfig.west);
			
			\draw [arrow] (compose.south) -- (trivial.north);
			\draw [arrow] (trivial.south) -- (overcomplete.north);
            \draw [arrow] (overcomplete.south) -- (symmetrize.north);
			\draw [arrow] (symmetrize.south) -- ++(0,-1cm) -- (removeodd.north east);
			\draw [arrow] (removeodd.south west) -- (preferred.north west);
            \draw [arrow] (removeodd.south east) -- (preferred.north east);
            \draw [arrow] (rewrite.south) -- ++(0,-0.3cm) -- (removeodd.north west);
            \draw [arrow] (preferred.south west) -- (relmat.north);

			\draw [arrow] (oprel.south) -- (rewrite.north);
			
			\draw [arrow] (relmat) -- (extrarel);
			\draw [arrow] (extrarel) -- (decompose);
			\draw [arrow] (decompose.east) -- (independent);
			
			\draw (oprelConfig.west) -- ++(- 0.3cm, 0) |- (oprel.west);
			
            \draw [arrow] (overcomplete.west) -| node[above, near start] {reference to} (oprel.north);
            
			\draw [discontinuous arrow] (symmetrize.south west) -- node[above, midway, sloped] {transformations} (rewrite.east);
			
			\draw (relmat) -- (system);
			\draw (extrarel) -- (contact);
			\draw (independent) -- (result);

		\end{tikzpicture}
	}
	\caption{Flow of a calculation with \minibar~\cite{MINIBAR}. The square boxes represent the main expressions and the diamond-shaped ones actions on them. Orange boxes indicate some of the specific configurations or results for our calculation. Lighter blue boxes indicate optional steps that do not change the underlying physical result. }
	\label{fig:scheme}
\end{figure}

The structure of the calculation with \minibar, which is essentially identical to the one in \textsc{Form}, is shown in Figure~\ref{fig:scheme}. The blue boxes and arrows show the steps for a full calculation of a minimal operator basis, whereas the orange boxes show the specific configuration for our calculation. The source code of the package and this specific example are accessible in the \minibar repository~\cite{MINIBAR}.

To start, the possible combinations of fields and derivatives with dimension $p^8$ and odd intrinsic parity are determined. Within each combination, all possible ways to order the fields, apply derivatives on them, and introduce flavour traces are \textit{written} down. For each of these, the indices on the fields (four contracted with the Levi-Civita symbol and up to two additional pairs of indices) are introduced in all possible permutations. Once all invariant structures are composed, we reduce this list employing some trivial index and trace relations, and store the dictionary to rewrite any term into one of the standard structures in the resulting overcomplete basis.

To find a minimal set of operators from the overcomplete basis we must explicitly derive all relations between them. In our case, there are up to seven types of identities that give rise to operator relations. These are defined in Section 3 of the main article~\cite{Bijnens:2023hyv}. To do this, \minibar is equipped with time-efficient functions to find relations from any identity, introduced as a replacement rule, whereas the operator relations from some generic identities can be produced with user-friendly functions that are integrated in the main \minibar code. These include the Cayley-Hamilton trace relations, for matrices of any dimension; the Schouten identity, that finds impossible antisymmetric index combinations; and the vanishing of total derivatives in the action. The operator relations are then simplified, writing each operator into the standard shape defined with the trivial relations. Next, the operators are put in linear combinations (monomials) that are Hermitian and $\mathcal{C}$-even, and these reordered to preferentially keep some of them after the final reduction. To that end, a matrix with sparse entries is built, capturing the linear system of equations defined by the operator relations, and an algorithm of Gaussian elimination particularly optimized for sparse matrices is employed to solve the system. To separate the unphysical contact operators from the rest of terms with mesons in the Lagrangian, the calculation was repeated in the LR basis using only external fields. The resulting monomials are then translated to the $u_\mu$ basis and employed to extend the relation matrix of the main calculation before solving the system. The minimal set of linearly-independent monomials is extracted.

\vspace{-0.3cm}
\section{Results and conclusions}
\vspace{-0.4cm}
We obtained the anomalous ChPT Lagrangian at NNLO, which is available in the supplementary material of the main article~\cite{Bijnens:2023hyv}, as well as in the \minibar repository~\cite{MINIBAR}. This is provided for 2, 3, and general number of flavours $N_f$, which differ due to the operator relations coming from the Cayley-Hamilton identities, that depend on the dimension of the $N_f \times N_f$ flavour matrices. We find 999 operators for a general $N_f$, 705 for $N_f=3$ and $211$ for $N_f=2$. Our result agrees with that of the Hilbert series, a group-theoretical method to determine the number of independent monomials without extracting the Lagrangian explicitly \cite{Graf:2020yxt}. In the appropriate limit, we also find agreement with a previous work on the non-linear sigma model~\cite{Kampf:2021jvf}. This result can also be used to compare with the Adler zero methods, see \textit{e.g.} Refs.~\cite{Low:2014nga,Low:2022iim}.
\vspace{-0.3cm}
\section*{Acknowledgements}
\vspace{-0.4cm}
J.~B.~and J.~R.~V.~are supported by the Swedish
Research Council, with grant contract numbers 2016-05996 and 2019-03779. 
N.~H.-T.~was first funded by the Swedish Research Council (project number 2021-06638) and now by the UK Research and Innovation, Engineering and Physical Sciences Research Council (grant number EP/X021971/1). 

\vspace{-0.3cm}
\bibliographystyle{JHEP}
\bibliography{refs}

\end{document}